# Vacancy Effects on Electric and Thermoelectric Properties of Zigzag Silicene Nanoribbons

Rui-Li An[1], Xue-Feng Wang[1,*], P. Vasilopoulos[2], Yu-Shen Liu[3], An-Bang Chen[1],

Yao-Jun Dong[1], and Ming-Xing Zhai[1]

[1] College of Physics, Optoelectronics and Energy, Soochow University, Suzhou 215006, China

[2] Concordia University, Department of Physics, Montreal, QC, Canada, H4B 1R6

[3] Jiangsu Laboratory of Advanced Functional materials, Changshu 215500, China

* E-mail: xf_wang1969@yahoo.com

We study the crystal reconstruction in the presence of monovacancies (MVs), divacancies (DVs) and linear vacancies (LVs) in a zigzag silicene nanoribbon (ZSiNR) with transversal symmetry. Their influence on the electric and thermoelectric properties is assessed by the density functional theory combined with the nonequilibrium Green's functions. In particular, we focus on the spin resolved conductance, magnetoresistance and current-voltage curves. A 5-atom-ring is formed in MVs, a 5-8-5 ring structure in DVs, and a 8-4-8-4 ring structure in LVs. The linear conductance becomes strongly spin dependent when the transversal symmetry is broken by vacancies especially if they are located on the ribbon's edges. The giant magnetoresistance can be smeared by asymmetric vacancies. Single spin negative differential resistance may appear in the presence of LVs and asymmetric MVs or DVs. A strong spin Seebeck effect is expected at room temperature in ZSiNRs with LVs.

1. **Introduction**

   Graphene nanoribbons (GNRs)[1,2] have attracted extensive attention for their outstanding electronic properties and their application potential in nanodevices[3-10]. In real devices on one hand defects are not avoidable but on the other hand they can be employed to improve the device performance. The effects of defects on the electronic properties of GNRs are interesting and have been studied intensively in the last years. It has been shown that the type and location of vacancies can greatly affect the geometric structure and electronic behavior of GNRs[11-19].

The silicon counterpart of graphene, silicene, has been predicted to have a similar electronic band structure as graphene[20,21]. Recently, silicene and silicene nanoribbons (SiNRs) were synthesized on various substrates[22-26]. Dirac cones similar to those in graphene were observed though their origin is still in controversy[27]. They are attracting increasing attention because they are compatible with current silicon technologies[28-30]. Different from graphene, silicene is not planar and a perpendicular electric field can open an energy band gap at the Fermi energy $E_F$. Furthermore, the spin-orbit interaction in silicene is about 1 meV and may be used to further manipulate the spin behavior in spintronic devices. Though the creation of free-standing silicene has yet not been reported and several claims are controversial, see, e.g., Ref.[30] for a recent review, SiNRs have been created and shown to have electronic properties more sensitive to the structural size and geometry than GNRs. The armchair-edge SiNRs (ASiNRs) are all semiconducting independent of their width because of their specific dimer edge, while zigzag-edge SiNRs (ZSiNRs) are metallic when the ribbons are wide[31]. Recently, the structural and electronic properties of SiNRs have been one of the research focuses[31-35]. Computational simulations have suggested that vacancies are easier to form and have similar geometries in silicene compared to those in graphene[36-39] though direct experimental evidence is yet to come out.

Thermoelectric effects convert heat energy into electric energy and vice versa. The spin Seebeck effect creates a spin voltage or a spin current from a temperature gradient in materials. In the last few years, spin voltages have been realized by the Seebeck effect in magnetic and nonmagnetic materials[40-46]. A spin Seebeck effect in zigzag GNRs has been predicted and is expected to be greatly enhanced by doping[8,47].

In this paper, using first-principles calculations, we focus on the effects of monovacancies (MVs), divacancies (DVs), and of linear defects on the structural and electronic properties of ZSiNRs. We consider MVs and DVs at various sites, as shown in Fig. 1, and evaluate the conductance and the current-voltage (I-V) characteristics in various electrode setups. In Sec. 2 we present the model and in Sec. 3 the results. Concluding remarks follow in Sec. 4.

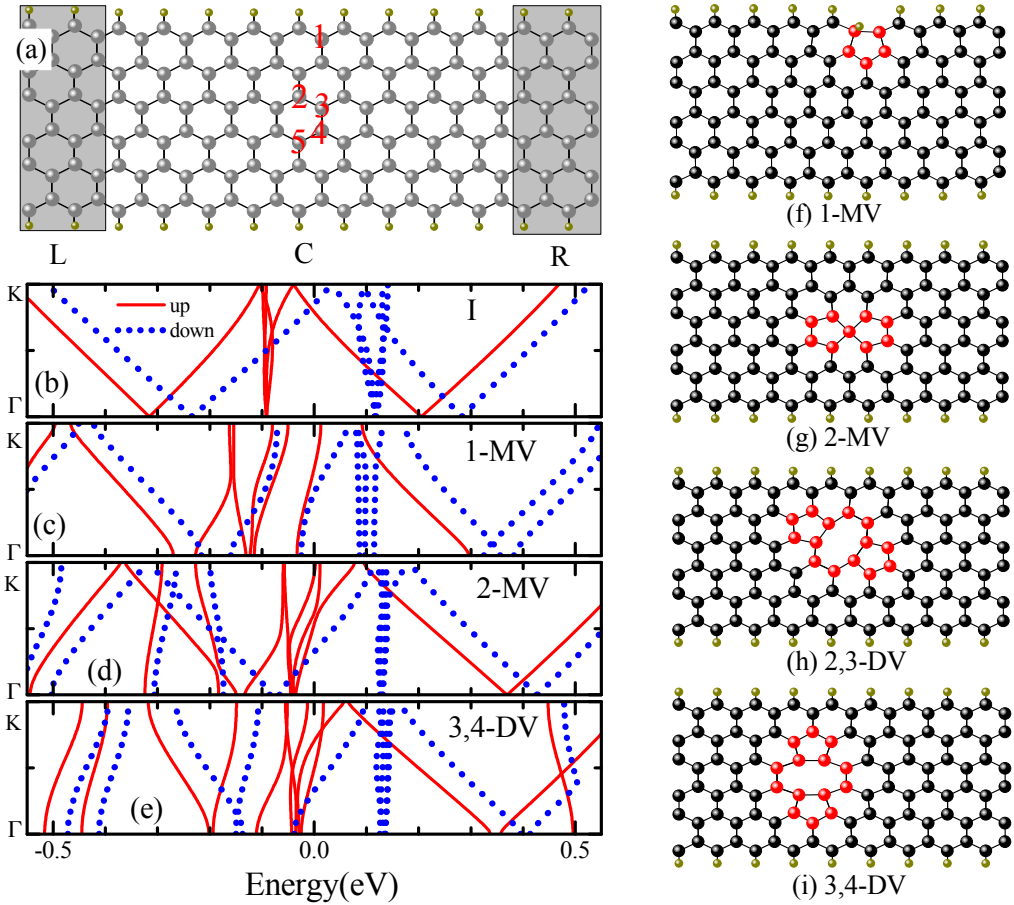

Fig. 1. (a) An ideal (I) ZSiNR, with a central region (C) of size 6 × 9, is connected to a left (L) and right (R) electrodes (grey areas). The big spheres are the Si atoms and the small ones the H atoms on the edges. Vacancy defects are formed by removing one or two atoms from sites 1-5. The optimized atomic structures of four typical defects, samples 1-MV, 2-MV, 2,3-DV, and 3,4-DV, are shown in panels (f)-(i), respectively. The electronic band structures of ferromagnetic (FM) samples I, 1-MV, 2-MV, and 3,4-DV are plotted in panels (b)-(e), respectively, for up-spins (solid curves) and down-spins (dotted curve). $E_F$ is assumed to be at zero energy ($\mu_0 = 0$).

## 2. Method and computational details

With the Atomistix Toolkit (ATK) package, we use the density functional theory (DFT) to calculate the electronic structures and the nonequilibrium Green's function (NEGF) method combined with it to simulate electron quantum transport with inelastic scatterings neglected. The exchange-correlation functional is treated in the spin dependent local density approximation with the Perdew-Zunger (SLDA-PZ) parameters. Note that use of the spin dependent generalized gradient approximation with the Perdew-Burke-Ernzerh (SGGA-PBE) parametrization leads to the same conclusion though some quantitative changes appear. The kinetic energy cutoffs for the wavefunction and charge are set to 40 Ry and 150 Ry, respectively. A supercell of ZSiNR, constructed by a 6×9 extension of the hexagonal unit cell of ideal silicene with lattice constant 3.87 Å as the central region shown in Fig. 1(a), is employed to carry out the structure optimization and energy band calculation. Here the width 6 means that the ribbon is composed of 6 zigzag Si chains and the length 9 indicates that there are 9 unit cells in the supercell. Change of the ribbon width will change slightly the energy gap near the Fermi energy and the distance between the edges and the vacancy in samples with defects, the results presented will vary quantitatively but the conclusion remains the same. The ZSiNR is transversally symmetric and more specifically rotationally symmetric about its central line[32,48]. First-principles calculations have shown that the edge atoms have dangling bonds and can be passivated by H atoms similar to the case in GNRs[49]. Clear ZSiNRs are nonmagnetic due to edge reconstruction and the passivated ones have spin polarized edge states. Here we assume that the Si atoms on the edges are passivated by H atoms. To avoid spurious interactions between ZSiNRs, a vacuum space of thickness 15 Å is inserted between neighboring layers and another one of thickness 36.1 Å between the edges on each layer. A 1×1×500 *k*-point grid in the Monkhorst-Pack scheme is employed to sample the Brillouin zone of the supercell. During the optimization, atoms are fully relaxed until residual forces on constituent atoms are smaller than 0.05 eV/Å.

The ideal ZSiNR without defect is denoted as sample I and the samples with vacancy defects are formed by removing one of two atoms at sites 1-5 as shown in Fig. 1(a). Four common types of native point vacancies, two MVs denoted as *n*-MV with *n* =1 or 2 labelling the vacancy site and two DVs, denoted as *m,n*-DV, are studied as representative vacancy defects in ZSiNRs. The relaxed structures of the samples, 1-MV, 2-MV, 2,3-DV, and 3,4-DV are shown in Fig. 1(f)-(i), respectively. Note that 2,4-DV and 2,5-DV have the same relaxed structure as 2,3-DV and 3,4-DV, respectively. In consistent with the results in the literature[36-38], the MV defects have 5-atom-ring structures and the DV defects have structures of 5-8-5 atomic rings. In addition, our calculations show that defects are easier to form on edges. Defect 1-MV has an energy 2.64eV lower than defect 2-MV, and defect 2,3-DV has an energy 0.22eV lower than defect 3,4-DV. Similar to zigzag GNRs[3,8,9], the two edges of ZSiNRs are spin polarized. The magnetization directions of the two edges are parallel to each other in the ferromagnetic (FM) state and antiparallel in the antiferromagnetic (AFM) state. Our calculation has shown that the AFM state is the ground state in the absence of an external magnetic field. The FM state can be reached with the application of such a field.

To have an idea how the vacancies affect the energy states in ZSiNRs, we calculate the energy bands of periodic bulk systems formed by repeating the supercell along the ribbon's direction. The energy band structures near $E_F = \mu_0 = 0$ for samples I, 1-MV, 2-MV, and 3,4-DV in their FM sates are plotted in Fig. 1(b)-(e), respectively. The edge states have energy bands with weak dispersion. In the ideal ZSiNR, the spin-up edge sates have energies about -0.10eV and the spin-down ones about 0.11eV. In sample 1-MV, with a MV on edge, the spin-up edge-state bands qcquire some extra dispersion and the spin-down ones shift to energy 0.10eV. In addition, some energy gaps appear near the $E_F$ for both spin-up and spin-down electrons. In sample 2-MV, with the MV located along the central line of the ribbon, the energy bands of the spin-up edge states shift to -0.04eV as shown in Fig. 1(d). No obvious energy gap appears near $E_F$. In samples with DVs, the vacancies are inside the ribbons and the edge states have similar energies as those in sample 2-MV as shown in Fig. 1(e) for sample 3,4-DV.

To study electron transport two-probe systems are established by connecting ZSiNRs, of size 6×9, with two semi-infinite ZSiNRs electrodes as illustrated in Fig.1 (a). For FM electrodes, their magnetizations can be parallel (FM-P) or antiparallel (FM-AP). When a voltage bias $V_b$ is applied between the electrodes we assume that $E_F$ shifted from its equilibrium value ($\mu_0$) to $\mu_L = \mu_0 - eV_b/2$ in the left (L) and $\mu_R = \mu_0 + eV_b/2$ in the right (R) electrode, respectively. The conductance $G_\sigma(E)$ for spin $\sigma$ at energy $E$ is given in terms of the electronic transmission $\tau_\sigma$ as

$$G_\sigma(E) = \frac{e^2}{h}\tau_\sigma(E) = \frac{e^2}{h}Tr[\Gamma_L G^R \Gamma_R G^A]_\sigma, \qquad (1)$$

where $G^R$ ($G^A$) is the retarded (advanced) Green's function and $\Gamma_L$ ($\Gamma_R$) the broadening matrix due to the L (R) electrode. The total conductance is defined as the sum of spin-up and spin-down components: $G_T = G_\uparrow + G_\downarrow$. The spin-dependent current $I_\sigma$ through the central region is evaluated by the Landauer-Büttiker formula

$$I_\sigma(V_b) = \frac{e}{h}\int_{-\infty}^{\infty}\tau_\sigma(E)[f(E-\mu_R) - f(E-\mu_L)]dE \qquad (2)$$

Here $f(x) = 1/[\exp(x/k_B T) + 1]$ is the Fermi distribution function with $k_B$ the Boltzmann constant, $T$ the temperature, and $\tau_\sigma$ the spin-dependent transmission.

The spin polarization in the linear-response regime is measured by

$$\zeta = \frac{\tau_\uparrow(\mu_0) - \tau_\downarrow(\mu_0)}{\tau_\uparrow(\mu_0) + \tau_\downarrow(\mu_0)} \qquad (3)$$

and the corresponding tunneling magnetoresistance (MR) of systems with FM electrodes by[50]

$$MR^{FM} = \frac{G_T^{FM-P} - G_T^{FM-AP}}{\text{Min}\{G_T^{FM-P}, G_T^{FM-AP}\}} \qquad (4)$$

where $G_T^{FM-P}$ and $G_T^{FM-AP}$ are the total conductances at $E = 0$ of the two-probe system in the P and AP configuration of the electrodes, respectively. By switching on and off the external magnetic field, we can drive the systems from the FM state to the AFM state and change its conductance. The corresponding magnetoresistance

between FM and AFM states can be defined as[51]

$$MR^B = \frac{G_T^{FM-P} - G_T^{AFM}}{\text{Min}\{G_T^{FM-P}, G_T^{AFM}\}}, \tag{5}$$

where $G_T^{AFM}$ is the total linear conductance of the system with AFM electrodes. Note that the spin orbit interaction (SOI) is not taken into account in the calculation. The SOI will open an energy gap about 1 meV in the FM state of ZSiNRs and can affect the linear conductance in real systems at low temperatures.

The spin-dependent Seebeck coefficient $S_\sigma$ describes the spin-dependent voltage bias $\Delta V_\sigma$ generated by a temperature difference $\Delta T$ between the two electrodes in an open circuit. In the linear response regime $S_\sigma$ is obtained by[52,53]

$$S_\sigma = -\lim_{\Delta T \to 0} \frac{\Delta V_\sigma}{\Delta T} = -\frac{1}{eT}\frac{K_{1\sigma}(\mu_0,T)}{K_{0\sigma}(\mu_0,T)}, \tag{6}$$

where $K_{\nu\sigma}(\mu_0,T) = \int dE\, [\partial f(E-\mu_0)/\partial E](E-\mu_0)^\nu \tau_\sigma(E)$ with $\nu = 0,1$. For low temperatures, $S_\sigma$ can be simplified as[8]

$$S_\sigma \approx -\frac{\pi^2 k_B^2 T}{3e} \dot{\tau}_\sigma(\varepsilon)/\tau_\sigma(\varepsilon)|_{\varepsilon=\mu_0} \tag{7}$$

Here we see that $S_\sigma$ is related not only to the value of $\tau_\sigma(\varepsilon)$ but also to its slope near $E_F$.

## 3. Results and discussion

### a. Conductance

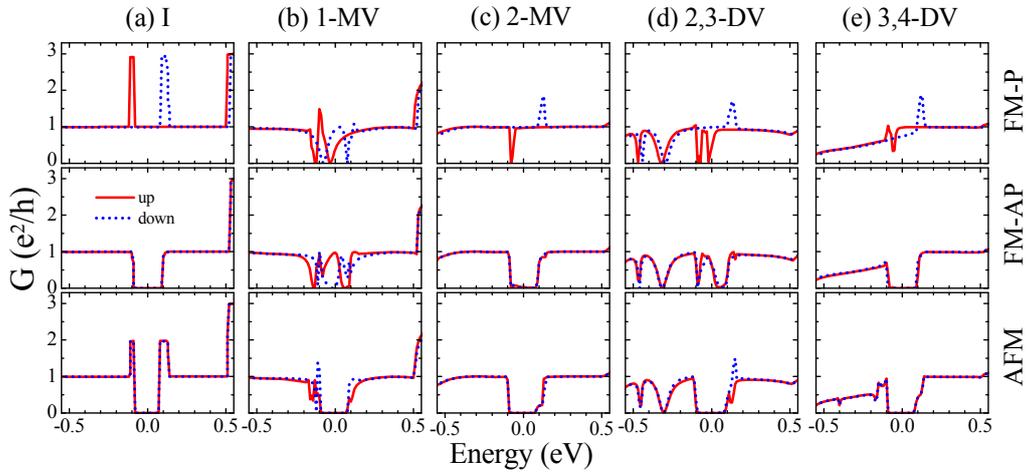

Fig. 2. Spin-up (solid curves) and spin-down (dotted curves) conductance

spectra of samples (a) I (ideal ZSiNR), (b) 1-MV, (c) 2-MV, (d) 2,3-DV, and (e) 3,4-DV in the FM-P (1st row), FM-AP (2nd row), and AFM (3rd row) configurations.

The conductance $G$ of sample I in the FM-P configuration is equal to one conductance quantum $G_0=e^2/h$ near $E_F$ except at the energies where the energy bands of the edge states become flat at E=-0.10eV for spin-up and at E=0.11eV for spin-down electrons as shown in the 1st row of Fig. 2(a). In the FM-AP configuration (2nd row) a conductance gap appears at $E=0$ due to the mismatch of wave functions in the two electrodes[54]. In the AFM state (3rd row) two conductance peaks exist besides the conductance gap at $E=0$ due to the flat band of the edge states besides the energy gap at $E=0$[48].

In the presence of a vacancy, fine structures in the conductance spectra appear due to the formation of bound states around the vacancy. They reflect the geometry characteristics of the vacancy and can be used as a finger print to identify the vacancy type. In sample 1-MV, the vacancy is located on the edge. It modifies greatly the edge states and the conductance mainly near $E=0$ as shown in Fig. 2(b). In addition, the 1-MV breaks the transversal symmetry of the system and the conductance spectra become spin dependent in the FM-AP and the AFM configurations. The wave function symmetry is also broken and the E=0 conductance gap in the FM-AP configuration disappears. In sample 2-MV, the vacancy is located in the central line of the ribbon and the atomic structure remains transversally symmetric. As a result, the conductance is only slightly modified as shown in Fig. 2(c).

In sample 2,3-DV, the vacancy is inside the ribbon but the transversal symmetry of the system is broken. Two extra Fano-type conductance dips, besides those near $E_F$, appear a little far away from $\mu_0$ in all configurations, at E=−0.3eV and −0.45eV, respectively. The band structure of the corresponding bulk system shows that two band gaps appear at these two energies. In the FM-P configuration, the conductance becomes strongly spin dependent near $\mu_0$ where two dips exist only in the spin-up conductance. The linear conductance in the FM-AP configuration becomes finite due to the change of geometric symmetry. In sample 3,4-DV the transversal symmetry of the system remains intact and its conductance spectrum is similar to that of sample 2-MV.

Table 1. Conductance and magnetoresistance in ideal and defected samples

| Sample | $G_T^{FM-P}$ (e²/h) | $G_T^{FM-AP}$ (e²/h) | $G_T^{AFM}$ (e²/h) | MR$^{FM}$(%) | MR$^B$ (%) |
|---|---|---|---|---|---|
| I | 2.00 | 2.84×10⁻⁵ | 8.98×10⁻¹⁰ | 7.03×10⁶ | 2.22×10¹¹ |
| 1-MV | 1.29 | 1.00 | 1.06×10⁻⁶ | 29.05 | 1.22×10⁸ |
| 2-MV | 1.96 | 0.0322 | 1.42×10⁻⁷ | 5.99×10³ | 1.38×10⁹ |
| 2,3-DV | 1.45 | 1.30 | 4.01×10⁻⁸ | 11.70 | 3.63×10⁹ |
| 3,4-DV | 1.72 | 2.45×10⁻³ | 1.16×10⁻⁷ | 7.01×10⁴ | 1.48×10⁹ |

The effects of vacancies on $G_T$ and MR$^{FM}$ of ZSiNRs are summarized in Table 1. $G_T^{FM-P} > e^2/h$ in all samples due to the matched wave function in the electrodes. $G_T^{FM-AP}$ is usually very small due to a mismatch of the wave functions but can be larger than $e^2/h$ in samples 1-MV and 2,3-DV with transversal symmetric vacancy. In these later cases, MR$^{FM}$ becomes quite large. $G_T^{AFM}$ is always small and MR$^B$ is extremely large since there is an energy gap at the Fermi energy in the AFM state of ZSiNRs. In Table 2 we present the spin polarization of the linear conductance. Sample 1-MV has a high spin polarization (95.3%) together with a large conductance in the FM-AP configuration.

Table 2. Spin polarization at the Fermi energy for different vacancy types and configurations.

| Sample | | I | 1-MV | 2-MV | 2,3-DV | 3,4-DV |
|---|---|---|---|---|---|---|
| Spin Polarization (%) | FM-P | 0.10 | 43.18 | 0.60 | 40.91 | 14.78 |
| | FM-AP | 3.63 | 95.30 | 52.60 | 0.85 | 17.60 |
| | AFM | 0.02 | 99.82 | 40.37 | 72.33 | 42.38 |

## b. Current-voltage characteristics

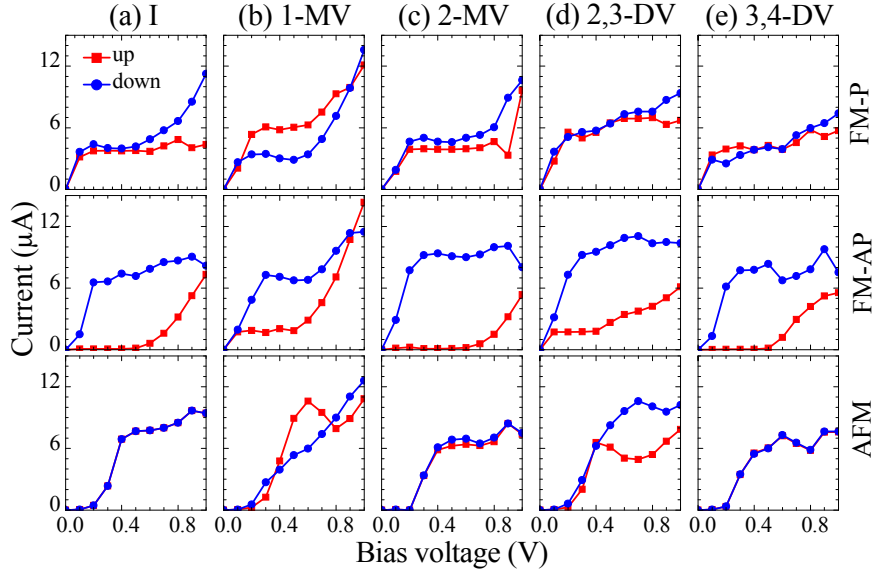

Fig. 3. The spin-up (solid curves) and spin-down (dotted curves) current-voltage characteristics of samples I, 1-MV, 2-MV, 2,3-DV, and 3,4-DV are plotted in panels (a)-(e), respectively, in the FM-P (1st row), FM-AP (2nd row), and AFM (3rd row) configurations.

In the FM-P configuration, as shown in the 1st-row panels of Fig. 3, the current increases with voltage rapidly under small bias and then remains in a level for a range of bias before increasing quickly again. The spin polarization is small under small or intermediate bias except for sample 1-MV where a vacancy is located on the edge. In the FM-AP configuration the energy bands of the L (R) electrode shift downward (upward). We define up (down) spin the majority (minority) spin in the left right in this work. Then the wave functions of the two electrodes remain mismatched (they overlap) inside the transport energy window, i.e. $\mu_L < E < \mu_R$, for up (down) spins under small and intermediate positive bias. The spin-up current remains low as shown in the 2nd-row panels of Fig. 3. On the contrary, the wave functions match gradually with the voltage. As a result, the spin-down current increases quickly with voltage and the spin polarization in the nonlinear transport regime is large.

In the AFM configuration, as shown in the bottom panels of Fig. 3, the current-voltage (I-V) curves show a semiconductor characteristic with a threshold voltage about 0.2 V for both spins (spin-up edge states on the lower edge). In transversally symmetric systems (samples I, 2-MV, and 3,4-DV) the electrons of both spins have similar behaviour and the spin polarization is negligible. In transversally asymmetric systems (samples 1-MV and 2,3-DV) a very interesting spin negative differential resistance (SNDR) develops. In sample 1-MV the SNDR phenomena occurs in the bias range 0.6-0.8 V. The spin-down current increases with the bias monotonicallly above the threshold. In contrast, the spin-up current reaches a maximum at about 0.6 V and then decreases significantly to a minimum at around 0.8 V before it increases again. In other words, the negative differential resistance (NDR) occurs in the range 0.6-0.8 V *only for spin-up electrons.* In sample 2,3-DV, the SNDR occurs in the bias range 0.4-0.9 V where the currents of opposite spins vary in opposite ways before they increase together again. The spin-up (spin-down) current decrease (increases) in the range 0.4-0.7 V and then increases (decreases) in the range 0.7-0.9 V.

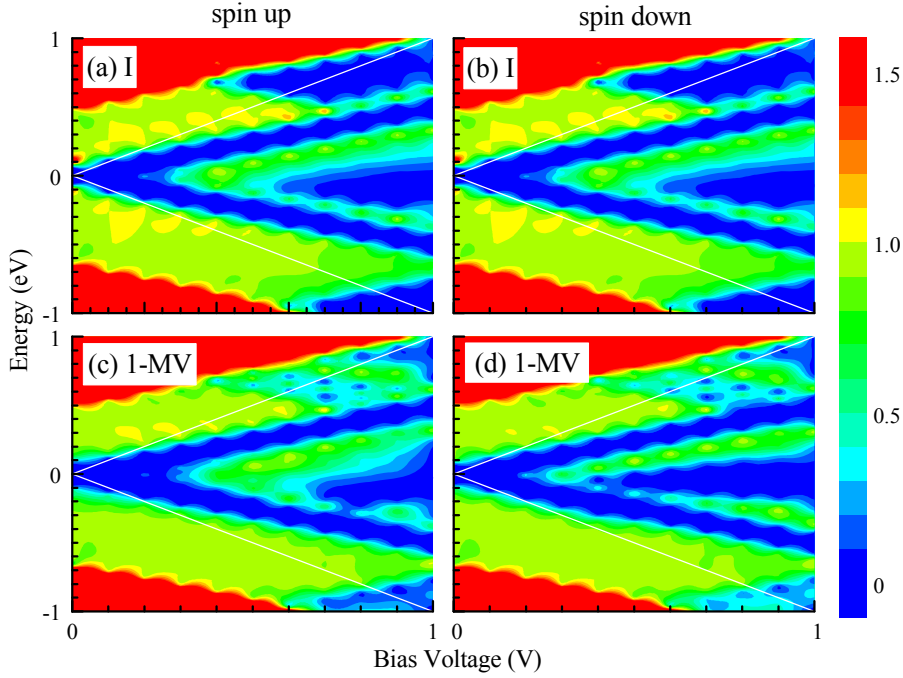

Fig.4. The spin-up, (a) and (c), and spin-down, (b) and (d), conductances ($e^2/h$) as functions of the electron energy and the applied bias voltage for samples I and 1-MV in the AFM configuration. The region between the two white solid lines is referred to as the transport window.

To better understand the SNDR effect that appeared in the AFM configuration of sample 1-MV, as illustrated in Fig. 3(b), we show 3D contour plots of the spin-up and spin-down conductances versus the applied bias voltage and the electron energy in Fig.4 for samples I in panels (a) and (b) and 1-MV in (c) and (d). The two white solid lines in each panel indicate the transport window in which the conductance contributes to the currents. There is a conductance gap of width 0.2 eV at $E_F$ under zero bias corresponding to the band gap of an ideal 6-ZSiNR. For bias higher than the threshold voltage 0.2 V, the transport window becomes wider than the conductance gap and a central conductance spectrum emerges at $E_F$. Both spin currents increase with voltage. In sample I, for both spins, an extra conductance gap appears in the central conductance spectrum at 0.5 V so the currents begin to increase at a much slower pace as shown in Fig. 3(a). In sample 1-MV, however, the electrons of opposite spins behavior in different way. For spin-down electrons confined mainly on the upper edge, where the vacancy is located, the extra conductance gap appears at a lower voltage 0.3 V so the spin-down current increases at an almost constant pace in a large range of bias. For spin-up electrons, on the contrary, the extra gap in the central conductance spectrum does not appear until the bias is higher than 0.6 V and the spin-up current increases quickly from 0.2 to 0.6 V as shown in Fig. 3(b). The extra spectrum gap becomes wide quickly once appeared so the current decreases with the voltage and the SNDR occurs.

*c. A line of vacancies*

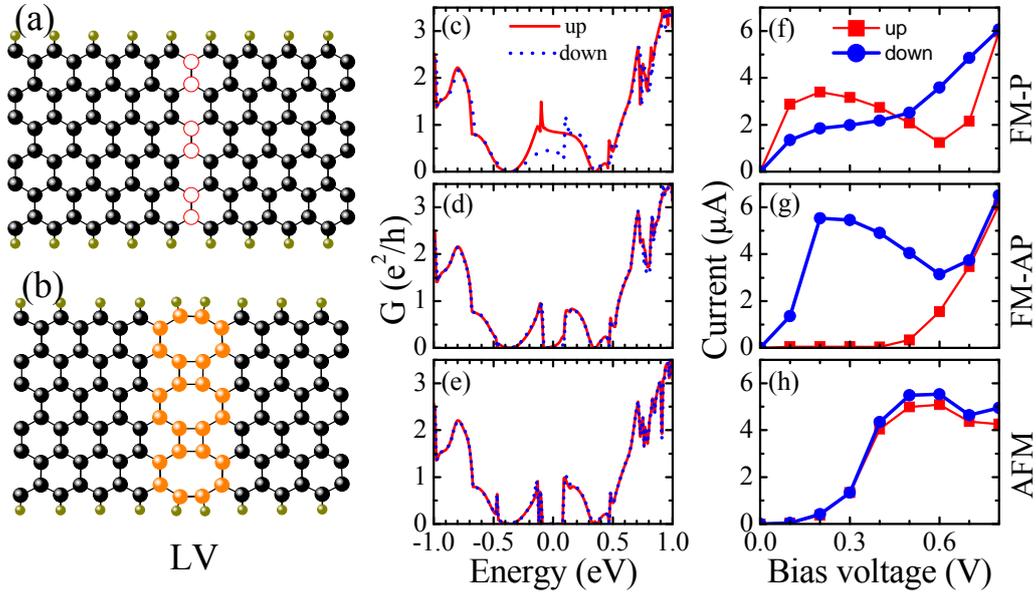

Fig. 5 (a) The vacant atoms of a line vacancy (LV) defect are marked by empty circles in an ideal system. (b) The optimized atomic structure when the Si atoms shown in (a) are removed. Si atoms near the removed ones are shown in orange, all other Si atoms are shown in black. The corresponding conductance spectrum is plotted in panels (c)-(e) in the FM-P, FM-AP, AFM configurations, respectively. Their current-voltage characteristics are shown, respectively, in panels (f)-(h).

We now consider transport when a line of Si atoms is vacant, such as the one formed by those Si atoms shown in red in Fig. 5(a). The corresponding conductance G as a function of the energy E, measured from $E_F$, is shown in Fig. 5(c)-(e), respectively, for the FM-P, FM-AP, and AFM configurations. As shown, after removal of the six Si atoms, the conductance spectra change greatly. The conductance plateaus at an integer number of $G_0$ disappear and two wide and dip conductance dips, one at -0.35 eV and another at 0.35 eV, appear. Since the structure remains transversally symmetric, the spin polarization is negligible in the FM-AP and AFM configurations under zero bias.

On the other hand, as shown in panels (f)-(h), the spin-up and spin-down currents, as functions of the voltage, are rather independent, i.e. the spin degeneracy is broken, in all configurations. Very interestingly, a SNDR occurs in the FM-P and FM-AP configurations. In the FM-P configuration the spin-down current increases monotonically with bias whereas the spin-up one has an obvious negative growth between 0.2 V and 0.6 V, and then it increases again. In the FM-AP configuration the spin-up current has a big threshold voltage (0.5 V) above which it increases with the bias monotonically. In contrast, the spin-down current reaches a maximum at about 0.2V and then decreases significantly to a minimum at around 0.6 V before it increases again. In other words, the NDR occurs between 0.2-0.6 V *only for spin-down electrons*. In the AFM configuration, there is a threshold voltage (0.2 V) for both spin currents. The I-V behaviours in the case of line defects are similar to the corresponding ones for MVs and DVs defects with transversal symmetry, i.e. 2-MV and 3,4-DV shown in Fig.3. However, the very interesting SNDR phenomenon appears in systems with line defects instead of those with MV and DV defects in the FM-P and FM-AP configurations.

### *d. Spin thermoelectric effects*

Thermoelectric properties, especially the Seebeck coefficient, significantly depend on the electronic band structure and are enhanced when $E_F$ is near the edges of a conductance gap. When the two spin channels are not mixed by spin-flip transitions and can be treated as independent in the whole system, the temperature gradient can lead not only to charge accumulation at the ends of an open system, but also to spin accumulation. In other words, the temperature gradient gives rise not only to an electric voltage, but also to a spin voltage. The spin-dependent Seebeck coefficients $S_\uparrow$ and $S_\downarrow$ calculated from Eq. (6) correspond to a spin-dependent voltage generated due to a temperature gradient. Generally, one can define the charge Seebeck coefficient as $S_C = (S_\uparrow + S_\downarrow)/2$ and the spin one as $S_S = (S_\uparrow - S_\downarrow)/2$[52,55].

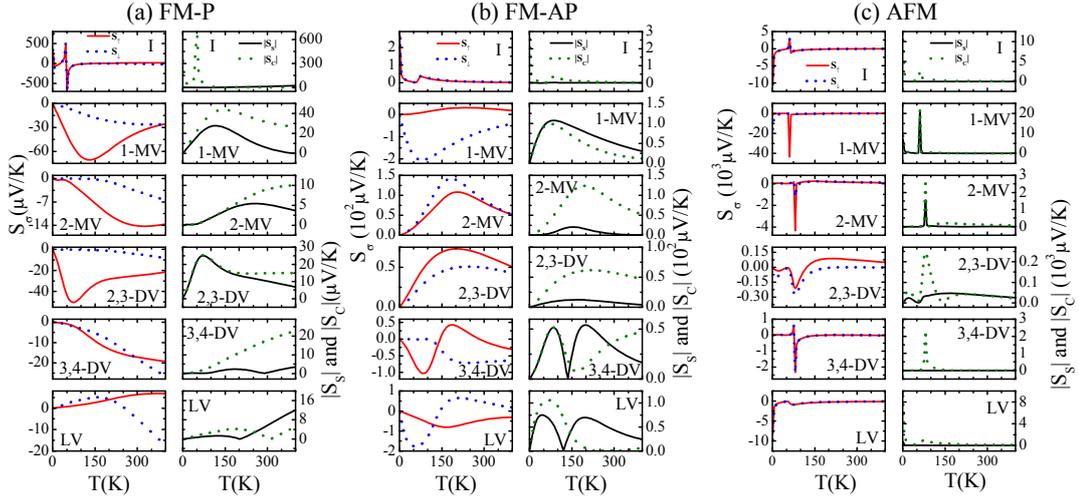

Fig. 6 Spin-dependent Seebeck coefficients $S_\sigma$ and the absolute values of spin ($S_S$) and charge ($S_C$) Seebeck coefficients as functions of temperature in (a) FM-P, (b) FM-AP and (c) AFM configurations for different samples.

In Fig.6, we present the spin-dependent Seebeck coefficients $S_\sigma$ (left column) and the absolute values of spin ($S_S$) and charge ($S_C$) Seebeck coefficients (right column) versus temperature in the FM-P (a), FM-AP (b), and AFM (c) configurations. The 6 rows are for samples marked, respectively, as I, 1-MV, 2-MV, 2,3-DV, 3,4-DV, and LV. In general, the sign of $S_\sigma$ indicates the type of devices: a negative (positive) Seebeck coefficient refers to an n-type (p-type) device. In the FM-P configuration, for MV and DV systems, $S_\uparrow$ and $S_\downarrow$ show a negative growth relation with the temperature at low temperatures, and may reach negative maximums. This suggests that ZSiNRs with MV and DV defects are n-type devices and the spin Seebeck coefficient is smaller than the charge one. In contrast, a ZSiNR with a linear defect has positive $S_\sigma$ at low temperature but $S_↓$ becomes negative at high temperatures as shown in the bottom panels of Fig. 6(a). In this case, a strong spin thermopower effect occurs with $|S_S| > |S_C|$ at room temperature. In the FM-AP setup, very interestingly, sample L has an almost zero $|S_C|$ at high temperatures with $|S_S| \gg |S_C|$. Samples 1-MV and 3,4-DV can have $|S_S| > |S_C|$ in some temperature region. In the AFM configuration, there is an energy gap of width 0.2 eV at $E_F$ in both electrodes and the linear conductance of all samples is very small. As a result, the Seebeck coefficients are also usually very small as shown in Fig. 6(c). However, the edges of the conductance gap, corresponding to the energy gap, at $\pm 0.1$ eV, are very sharp and result in a sharp increase of the Seebeck coefficients especially in samples with MV and DV defects at temperatures about 50K.

**4. Conclusions**

We have evaluated the spin-dependent conductance, current-voltage curves, and the Seebeck coefficients in zigzag silicene nanoribbons (ZSiNRs) employing the density functional theory combined with the Green's function method. Two-probe systems were studied in their ferromagnetic states, with various vacancy defects in the central region, and their electrodes in a parallel (FM-P) or antiparallel (FM-AP) configuration, as well as in their antiferromagnetic states (AFM). We discussed the effects of monovacancies (MVs), divacancies (DVs), and linear vacancies (LVs) on their properties. The linear conductance becomes strongly spin dependent if the transversal symmetry of the ZSiNRs is broken by the vacancies. A giant magnetoresistance between the FM and AFM states occurs in all cases but that between the FM-P and FM-AP configurations can be smeared out by asymmetric vacancies. A single-spin negative differential resistance is predicted in systems of AFM electrodes with asymmetric MVs or DVs. It is also predicted in systems of FM electrodes with LVs. The spin Seebeck coefficient can be larger than the charge one at room temperature in ZSiNRs with LVs. A spin NDR is predicted in ZSiNRs with edge defects in a AFM configuration and with line defects in a FM-P configuration.

**Acknowledgments**

This work was supported by the National Natural Science Foundation in China (Grant Nos. 11074182, 91121021, 11247028, and 61106126) and by the Canadian NSERC (Grant No. OGP0121756).

**Table of Contents**

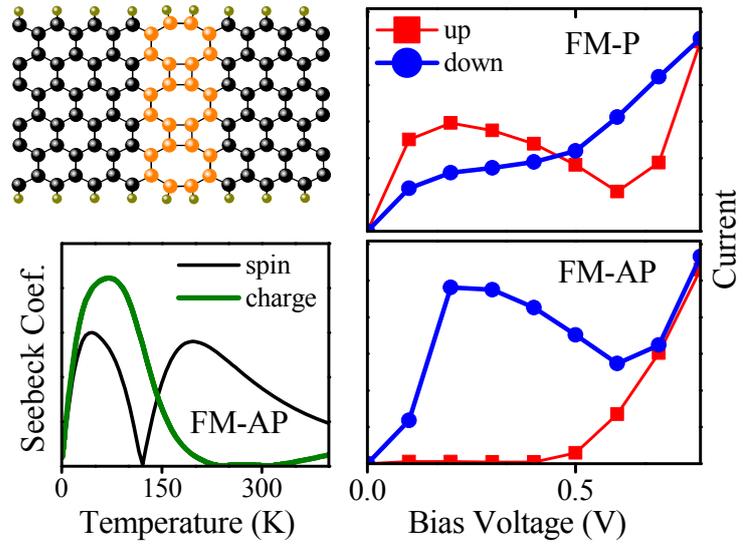